\documentclass[aps,pra,twocolumn,superscriptaddress]{revtex4-1}

\usepackage[table,x11names,dvipsnames,table]{xcolor}
\usepackage[linesnumbered,algoruled,boxed,lined]{algorithm2e}
\usepackage{amsfonts}
\usepackage{amsmath,tabu}    
\usepackage[title,toc,page]{appendix}
\usepackage{array}
\usepackage{booktabs}

\usepackage{color}      
\usepackage{colortbl}
\usepackage{cprotect}
\usepackage{dsfont}

\usepackage{fancyhdr}
\usepackage{fancyvrb} 
\usepackage{float}

\usepackage{graphicx}
\usepackage{hyperref}

\usepackage{listings}
\usepackage{pgfplots} 
\usepackage{tasks}
\usepackage{tikz}
\usetikzlibrary{angles, quotes, matrix, backgrounds, decorations.pathreplacing, positioning, arrows.meta}
\usepackage{todonotes}
\usepackage{verbatim}   

\newcolumntype{L}[1]{>{\raggedright\let\newline\\\arraybackslash\hspace{0pt}}m{#1}}
\newcolumntype{C}[1]{>{\centering\let\newline\\\arraybackslash\hspace{0pt}}m{#1}}
\newcolumntype{R}[1]{>{\raggedleft\let\newline\\\arraybackslash\hspace{0pt}}m{#1}}

\makeatletter

    \def\CT@@do@color{%
      \global\let\CT@do@color\relax
            \@tempdima\wd\z@
            \advance\@tempdima\@tempdimb
            \advance\@tempdima\@tempdimc
    \advance\@tempdimb\tabcolsep
    \advance\@tempdimc\tabcolsep
    \advance\@tempdima2\tabcolsep
            \kern-\@tempdimb
            \leaders\vrule
                    \hskip\@tempdima\@plus  1fill
            \kern-\@tempdimc
            \hskip-\wd\z@ \@plus -1fill }
    \makeatother

\newcommand{\lee}[1]{\textcolor{red}{#1}}

\begin{document}

\title{A hybrid classical-quantum workflow for natural language processing}

\author{Lee J. O'Riordan}
\email{Corresponding author: \\lee.oriordan@ichec.ie}
\affiliation{Irish Centre for High-End Computing, Dublin, Ireland.}%
\affiliation{National University of Ireland, Galway, Ireland.}

\author{Myles Doyle}%
\affiliation{Irish Centre for High-End Computing, Dublin, Ireland.}%
\affiliation{National University of Ireland, Galway, Ireland.}

\author{Fabio Baruffa}
\affiliation{Intel Deutschland GmbH, Feldkirchen, Germany.}%

\author{Venkatesh Kannan}
\affiliation{Irish Centre for High-End Computing, Dublin, Ireland.}%
\affiliation{National University of Ireland, Galway, Ireland.}

\date{\today}


\begin{abstract}
Natural language processing (NLP) problems are ubiquitous in classical computing, where they often require significant computational resources to infer sentence meanings. With the appearance of quantum computing hardware and simulators, it is worth developing methods to examine such problems on these platforms. In this manuscript we demonstrate the use of quantum computing models to perform NLP tasks, where we represent corpus meanings, and perform comparisons between sentences of a given structure. We develop a hybrid workflow for representing small and large scale corpus data sets to be encoded, processed, and decoded using a quantum circuit model. In addition, we provide our results showing the efficacy of the method, and release our developed toolkit as an open software suite.
\end{abstract}


\maketitle

\section{Introduction}
Natural language processing (NLP) is an active area of both theoretical and applied research, and covers a wide variety of topics from computer science, software engineering, and linguistics, amongst others. NLP is often used to perform tasks such as machine translation, sentiment analysis, relationship extraction, word sense disambiguation and automatic summary generation~\cite{nlp_review}. Most traditional NLP algorithms for these problems are defined to operate over strings of words, and are commonly referred to as the ``bag of words'' approach~\cite{Harris_BOW_54}. The challenge, and thus limitation, of this approach is that the algorithms analyse sentences in a corpus based on meanings of the component words, and lack information from the grammatical rules and nuances of the language. Consequently, the qualities of results from these traditional algorithms are often unsatisfactory when the complexity of the problem increases.

On the other hand, an alternate approach called ``compositional semantics'' incorporates the grammatical structure of sentences from a given language into the analysis algorithms. Compositional semantics algorithms include the information flows between words in a sentence to determine the meaning of the whole sentence~\cite{zadrozny-1992-compositional}. One such model in this class is ``(categorical) distributional compositional semantics'', known as DisCoCat~\cite{Coecke_Sadrzadeh_Clark_2010,Zeng_Coecke_2016,Coecke2019TheMO}, which is based on tensor product composition to give a grammatically informed algorithm that computes the meaning of sentences and phrases. This algorithm has been noted to potentially offer improvements to the quality of results, particularly for more complex sentences, in terms of memory and computational requirements. However, the main challenge in its implementation is the need for large classical computational resources.

With the advent of quantum computer programming environments, both simulated and physical, a question may be whether one can exploit the available Hilbert space of such systems to carry out NLP tasks. The DisCoCat methods have a natural extension to a quantum mechanical representation, allowing for a problem to be mapped directly to this formalism~\cite{Zeng_Coecke_2016}. Using an oracle-based access pattern, one can bound the number of accesses required to create the appropriate states for use by the DisCoCat methods~\cite{Wiebe_Kapoor_Svore_2014}. Though, this requires the use of a quantum random access memory, or qRAM~\cite{Giovannetti_Lloyd_Maccone_2008, Arunachalam_Gheorghiu_Jochym-OConnor_Mosca_Srinivasan_2015}. Currently, qRAM remains unrealised, and expectations are that the resources necessary to realise are as challenging as a fault tolerant quantum computer~\cite{qram_resource_2020}. As such, it can be useful to examine scenarios where qRAM is not part of the architectural design of the quantum circuit. This will allow us to examine proof-of-concept methods to explore and develop use-cases later improved by its existence.

In this paper we examine the process for mapping a corpus to a quantum circuit model, and use the encoded meaning-space of the corpus to represent fundamental sentence meanings. With this representation we can examine the mapping of sentences to the encoding space, and additionally compare sentences with overlapping meaning-spaces. We follow a DisCoCat-inspired formalism to define sentence meaning and similarity based upon a given compositional sentence structure, and relationships between sentence tokens determined using a distributional method of token adjacency.

This paper will be laid out as follows:
Section~\ref{sec:qnlp} will give an introduction to NLP, the application of quantum models to NLP, and discuss the encoding strategy for a quantum circuit model. Section~\ref{sec:methods} will discuss the preparation methods required to enable quantum-assisted encoding and processing of the text data. Section~\ref{sec:results} will demonstrate the proposed methods using our quantum NLP software toolkit~\cite{QNLP_REPO} sitting atop Intel Quantum Simulator (IQS)~\cite{qhipster2020}. For this we showcase the methods, and compare results for corpora of different sizes and complexity. Finally, we conclude in Section~\ref{sec:conclusions}.


\section{\label{sec:qnlp} NLP methods}
One of the main concerns of NLP methods is the extraction of information from a body of text, wherein the data is not explicitly structured; generally, the text is meant for human, rather than machine, consumption~\cite{NISBET2009119}. As such, explicit methods to infer meaning and understand a body of text are required to encode such data in a computational model.

Word embedding models, such as \texttt{word2vec}, have grown in popularity due to their success in representing and comparing data using vectors of real numbers~\cite{MikolovTomas2013EEoW}. Additionally, libraries and toolkits such as NLTK~\cite{nltk2009} and spaCy~\cite{spacy2} offer community developed models and generally incorporate the latest research methods for NLP. The use of quantum mechanical effects for embedding and retrieving information in NLP has seen much interest in recent years~\cite{Semantic_Composition_Meas2014, AertsDiederik2014MaQI, Wang:2019:DCM:3331184.3331412, Jaiswal2018QuantumlikeGO, TiwariPrayag2018MCMI, Wang2019, wiebe_quantum_2019}. 

An approach that aims to overcome the ambiguity offered by traditional NLP methods, such as the bag-of-words model is the categorical distributional-compositional (DisCoCat) model~\cite{Coecke_Sadrzadeh_Clark_2010, Zeng_Coecke_2016}. This method incorporates semantic structure, where sentences are constructed through a natural tensoring of individual component words following a set of rules determined from category theory. These rule-sets for which sentence structures may be composed are largely based on the framework of pre-group grammars~\cite{Lambek2008}. 

The DisCoCat approach offers a means to employ grammatical structure of sentences with token relationships in these sentences. Words that appear closer in texts are more likely to be related, and sentence structures can be determined using pre-group methods. These methods can easily be represented in a diagrammatic form, and allow for a natural extension to quantum state representation~\cite{Coecke2019TheMO}. This diagrammatic form, akin to a tensor network, allows for calculating the similarity between other sentences. This similarity measure assumes an encoded quantum state representing the structure of the given corpus, and an appropriately prepared test state to compare with. This alludes to a tensor-contraction approach to perform the evaluation.

While this approach has advantages in terms of accuracy and generalisation to complex sentence structures, state preparation is something we must consider. Given the current lack of qRAM, the specified access bounds are unrealised~\cite{Wiebe_Kapoor_Svore_2014}, and so it is worth considering state preparation as part of the process. Ensuring an efficient preparation approach will also be important to enable processing on a scale to rival that of traditional high-performance computing NLP methods. 

As such, we aim to provide a simplified model, framework and hybrid workflow for representing textual data using a quantum circuit model. We draw inspiration from the DisCoCat model to preprocess our data to a structure easily implementable on a quantum computer. We consider simple sentences of the form ``noun - verb - noun'' to demonstrate this approach. All quantum circuit simulations and preprocessing is performed by our quantum NLP toolkit (QNLP), sitting atop the Intel Quantum Simulator (formerly qHiPSTER) to handle the distributed high-performance quantum circuit workloads~\cite{qhipster2016, qhipster2020}. We release our QNLP toolkit as an open source (Apache 2.0) project, and have made it available on GitHub~\cite{QNLP_REPO}.

\section{\label{sec:methods}Methods}
\subsection{Representing meaning in quantum states}\label{subsec:state_meaning}

In this section, we discuss the implementation of the algorithms required to enable encoding, processing, and decoding of our data. We consider a simplified restricted example of the sentence structure ``noun-verb-noun'' as the representative encoding format. To represent sentence meanings using this workflow, we must first consider several steps to prepare our corpus data set for analysis:
\begin{enumerate}
    \item Data must be pre-processed to tag tokens with the appropriate grammatical type; stop-words (e.g. ``the'', ``a'', ``at'', etc.) and problematic (e.g. non-alphanumeric) characters should be cleaned from the text to ensure accurate tagging, wherein type information is associated with each word.
    \item The pre-processed data must be represented in an accessible/addressable (classical) memory medium. 
    \item There must be a bijective mapping between the pre-processed data and the quantum circuit representation to allow both encoding and decoding.
\end{enumerate}

Assuming an appropriately prepared dataset, the encoding of classical data into a quantum system can be mapped to two different approaches: state (digital), or amplitude (analogue) encoding~\cite{Schuld_2017,Mitarai_Kitagawa_Fujii_2019}. We aim to operate in a mixed-mode approach: encoding and representing corpus data using state methods, then representing and comparing test sentence data through amplitude adjustment, measurement, and overlap.

Our approach to encoding data starts with defining a fundamental language (basis) token set for each representative token meaning space (subject nouns, verbs, object nouns). The notion of similarity, and hence orthogonality, with language can be a difficult problem. Do we consider the words ``stand'' and ``sit'' to be completely opposite, or are they similar because of the type of action taken? For this work, we let the degree of `closeness' be determined by the distributional nature of the terms in the corpus; words further apart in the corpus are more likely to be opposite.

To efficiently encode the corpus data, we decide to represent the corpus in terms of the $n$ most fundamentally common tokens in each meaning space. This draws similarity with the use of a word embedding model to represent a larger space of tokens in terms of related meanings in a smaller space~\cite{levy-goldberg-2014-linguistic,socher-etal-2013-parsing,Mikolov2013_embedding}. This is necessary as representing each token in the corpus matching the sentence structure type can create a much larger meaning space than is currently representable, given realistic simulation constraints. However, one can note as we increase the limit of fundamental tokens in our basis, we tend to the full representative meaning model.

Taking inspiration from the above methods, we implement an encoding strategy that given the basis tokens, maps the remaining non-basis tokens to these, given some distance cut-off in the corpus. A generalised representation of each token $t_i$, in their respective meaning space $\mathbf{m}$ would be defined as 
\begin{equation}\label{eqn:map_tokens}
    t_i = \displaystyle\sum\limits_{j}^{n} d_{i,j} m_j
\end{equation}
where $d_{i,j}$ defines the distance between the base token $m_j$ and non-base $t_i$. As such, we obtain a linear combination of the base tokens with representative weights to describe the mapped tokens. 

We have identified the following key steps to effectively pre-process data for encoding:

\begin{enumerate}
    \item Tokenise the corpus and record position of occurrence in the text.
    \item Tag tokens with the appropriate meaning space type (e.g. noun, verb, stop-word, etc.)
    \item Separate tokens into noun and verb datasets.
    \item Define basis tokens in each set as the $N_\textrm{nouns}$ and $N_\textrm{verbs}$ most frequently occurring tokens.
    \item Map basis tokens in each respective space to a fully connected graph, with edge weights defined by the minimum distance between each other basis token.
    \item Calculate the shortest Hamiltonian cycle for the above graph. The token order within the cycle is reflective of the tokens' separation within the text, and a measure of their similarity.
    \item Map the basis tokens to binary strings, using a given encoding scheme.
    \item Project composite tokens (i.e. non-basis tokens) onto the basis tokens set using representation cut-off distances for similarity, $W_\textrm{nouns}$ and $W_\textrm{verbs}$.
    \item Form sentences by matching composite \textsc{noun-verb-noun} tokens using relative distances and a \textsc{noun-verb} distance cut-off, $W_\textrm{nv}$.
\end{enumerate}

After conducting the pre-processing steps, the corpus is represented as a series of binary strings of basis tokens. At this stage the corpus is considered prepared and can be encoded into a quantum register.


\subsection{Token encoding}\label{subsec:basis_words}

To ensure the mapping of the basis words to encoding pattern is reflective of the underlying distributional relationships between words in the corpus, it is necessary to choose an encoding scheme such that the inter-token relationships are preserved. While many more complex schemes can give insightful relationships, we choose a cyclical encoding scheme where the Hamming distance, $d_H$, between each bit-string is equal to the distance between the bit-strings in the data set. For 2 and 4-qubit registers respectively, this would equate to the patterns
\begin{align*}
    p_2 &= [00,\, 01,\, 11,\, 10], \\
    p_4 &= [0000,\, 0001,\, 0011,\, 0111,\, 1111,\, 1110,\, 1100,\, 1000 ],
\end{align*}
of which the $i = [1,2n]$ range of indexed (base-10) elements of $p_n$ can be generated iteratively by 
\begin{equation}\label{eqn:cyclic_encoding}
    p_{n}(i+1)=\begin{cases}
    2p_{n}(i) + 1, & \text{if $i \leq n + 1$},\\
    p_{n}(n+1) - p_{n}(i-n), & \text{if $n+1 < i \leq 2n$},\\
    \textrm{undefined} & \textrm{otherwise},
  \end{cases}
\end{equation}
given $p_n(1) = 0$. For the simple 2-bit pattern, this equates to a Gray code mapping, but differs for larger register sizes. With this encoding scheme, we can show that the Hamming distances between each pattern and others in the set have a well-defined position-to-distance relationship.

As an example, let us consider a 4-element basis of tokens given by $b=$\{\verb+up+, \verb+down+, \verb+left+, \verb+right+\}. We define \verb+up+ and \verb+down+ as opposites, and so should preserve the largest Hamming distance between them. This requires mapping the tokens to either 00,11, or to 10,01 for these pairs. Similarly, we find the same procedure with the remaining tokens. In this instance, we have mapped the tokens as
\begin{align*}
    \textrm{\verb~up~} \rightarrow 00;~ \textrm{\verb~down~} \rightarrow 11; ~\textrm{\verb~left~} \rightarrow 01; ~\textrm{\verb~right~} \rightarrow 10;
\end{align*}
which preserves the relationships we have discussed earlier in \ref{subsec:state_meaning}.

Once again, it is worth noting that the notion of similarity is complex when considering words, as such is the concept of orthogonality. It may be argued that the \verb+up+-\verb+down+ relationship has more similarities than, say, a \verb+left+-\verb+down+ relationship, but for the purpose of our example this definition is sufficient. Care ought to be taken into defining inter-token relationships, requiring some domain expertise of the problem being investigated. The choice of inter-token relationship taken during preparation will influence the subsequent token mappings determined later in the process.

For our work we have deemed it sufficient to define these similarities by distance between the tokens in a text; larger distances between tokens defining a larger respective Hamming distance, and smaller distances a smaller one. We can similarly extend this method to larger datasets, though the ordering problem requires a more automated approach. 
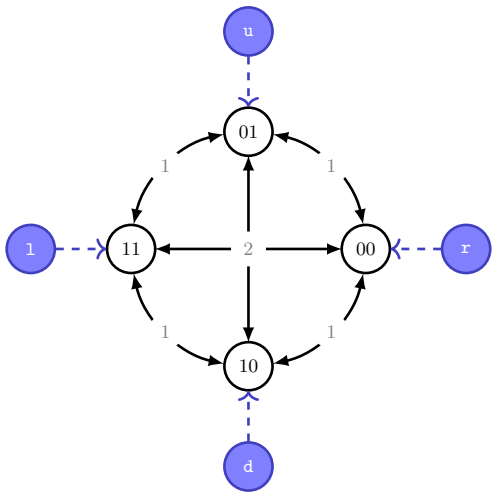
\begin{figure}
    \begin{tikzpicture}[scale=1.2, every node/.style={scale=0.8}]
        \def \radius {1.3cm}
        \def \margin {12}
        
        \node[draw, circle, line width=1pt, minimum size=0.8cm] at ({0}:\radius) { 00 };
        \node[draw, circle, line width=1pt, minimum size=0.8cm] at ({90}:\radius) { 01 };
        \node[draw, circle, line width=1pt, minimum size=0.8cm] at ({180}:\radius) { 11 };
        \node[draw, circle, line width=1pt, minimum size=0.8cm] at ({270}:\radius) { 10 };

        \node[draw, circle, line width=1pt, right=1cm of {0}:\radius, color=gray!50!blue, fill=white!50!blue, minimum size=0.8cm] at ({0}:\radius) (right) { \color{white} \verb+r+};
        \node[draw, circle, line width=1pt, left=1cm of {180}:\radius, color=gray!50!blue, fill=white!50!blue, minimum size=0.8cm] at ({180}:\radius) (left) { \color{white} \verb+l+};
        \node[draw, circle, line width=1pt, above=1cm of {90}:\radius, color=gray!50!blue, fill=white!50!blue, minimum size=0.8cm] at ({90}:\radius) (up) {\color{white} \verb+u+};
        \node[draw, circle, line width=1pt, below=1cm of {270}:\radius, color=gray!50!blue, fill=white!50!blue, minimum size=0.8cm] at ({270}:\radius) (down) {\color{white} \verb+d+};

        \draw[dashed, -{Computer Modern Rightarrow}, >=latex, line width=1pt, color=gray!50!blue,] (right) -- ({0}:\radius + 0.65*\margin);
        \draw[dashed, -{Computer Modern Rightarrow}, >=latex, line width=1pt, color=gray!50!blue,] (left) -- ({180}:\radius + 0.65*\margin);
        \draw[dashed, -{Computer Modern Rightarrow}, >=latex, line width=1pt, color=gray!50!blue,] (up) -- ({90}:\radius + 0.65*\margin);
        \draw[dashed, -{Computer Modern Rightarrow}, >=latex, line width=1pt, color=gray!50!blue,] (down) -- ({270}:\radius + 0.65*\margin);
    
        \draw[<->, >=latex, line width=1pt] ({0   +\margin}:\radius) arc ({0    +\margin}:{90    -\margin}:\radius);
        \draw[<->, >=latex, line width=1pt] ({90  +\margin}:\radius) arc ({90   +\margin}:{180   -\margin}:\radius);
        \draw[<->, >=latex, line width=1pt] ({180 +\margin}:\radius) arc ({180  +\margin}:{270   -\margin}:\radius);
        \draw[<->, >=latex, line width=1pt] ({270 +\margin}:\radius) arc ({270  +\margin}:{360     -\margin}:\radius);
    
        \draw[<->, >=latex, line width=1pt] ({0}:\radius*0.8) to ({180}:\radius*0.8) {};
        \draw[<->, >=latex, line width=1pt] ({90}:\radius*0.8) to ({270}:\radius*0.8) {};  
        \node[draw, circle, fill=white, color=white] at (0,0) {\color{gray} $2$ };
        \node[draw, circle, fill=white, color=white] at ({0   +45}:\radius) {\color{gray} $1$ };
        \node[draw, circle, fill=white, color=white] at ({90  +45}:\radius) {\color{gray} $1$ };
        \node[draw, circle, fill=white, color=white] at ({180 +45}:\radius) {\color{gray} $1$ };
        \node[draw, circle, fill=white, color=white] at ({270 +45}:\radius) {\color{gray} $1$ };
    \end{tikzpicture}
    \caption{Graph showing the mapping of tokens (blue) to bit-strings (white). Edge weights between the bit-strings represent the Hamming distances, $d_H$ between connected nodes. By mapping the tokens to the appropriate basis bit-string we can use the Hamming distances to represent differences between tokens.}\label{fig:tikz_4node_graph}
\end{figure}

For the 4-qubit encoding scheme, we must define a strategy to map the tokens to a fully connected graph, where again the respective positions of the bit-strings reflect the Hamming distance between them, as shown in Fig.~\ref{fig:tikz_4node_graph}. To effectively map tokens to these bit-strings, we use the following procedure:

\begin{enumerate}
    \item Given the chosen basis tokens, and their positions in the text, create a graph where each basis token is a single node.
    \item Calculate the distances between all token positions given a pairing of each token with the others, for all  $(n^2-n)/2$ pairings.
    \item For simplicity, choose the minimum distances between each of the pairings, and create edges with this as the given weight. As an aside, alternative methods can also be used, such as mean, median, etc.
    \item With the given fully-connected graph, find the minimum Hamiltonian cycle, and use the returned ordering to map the tokens onto the bit-strings.
\end{enumerate}

For the calculated minimum Hamiltonian cycle, the relationships between each of the tokens will be preserved, and can effectively be mapped onto the bit-string encoding scheme. It can be noted that alternative encoding schemes and distance orderings could potentially be investigated, but will remain beyond the scope of this current work. For our purposes we make use of the \verb+networkx+ package for the finding the minimum Hamiltonian cycle~\cite{networkx}.


\subsection{Methods for quantum state encoding}
To simplify our encoding procedure, we can assume a binary representation of distance for eq.~\eqref{eqn:map_tokens}, wherein all tokens within the given cutoff are equally weighted. This allows us to encode the states as an equal-weighted superposition, and is easily implemented as a quantum circuit~\cite{Trugenberger_2001,Trugenberger_2002}.

For notational simplicity, we define the following mappings:
\begin{equation*}
    \textrm{X}_{a}: \vert a \rangle \rightarrow \vert \lnot a \rangle,
\end{equation*}
\begin{equation*}
    \textrm{CX}_{a,b}: \vert a \rangle \vert b \rangle \rightarrow \vert a \rangle \vert a \oplus b \rangle
\end{equation*}
\begin{align*}
    \textrm{nCX}_{a_{1},\dots a_{n} ,b} : &\vert a_1 \rangle \dots\vert a_{n} \rangle \vert b \rangle \rightarrow \\ &\vert a_1 \rangle \dots \vert a_{n} \rangle \vert b \oplus (a_{1} \wedge \dots \wedge a_{n}) \rangle  
\end{align*}
where $\vert a \rangle$ and $\vert b \rangle$ are computational basis states, X is the Pauli-X ($\sigma_x$) gate, CX and nCX are the controlled X, and $n$-controlled NOT (nCX) operations, respectively. Additionally, we may define controlled operations using any arbitrary unitary gate using a similar construction of the above.

The goal of this algorithm is to encode a set of bit-strings representing our token meaning-space as an equal weighted superposition state. For a set of $N$ unique binary patterns $p^{(i)} = \{p_1^{(i)}, \dots, p_n^{(i)} \}$ each of length $n$ for $i=1,\dots,N$, we require three registers of qubits; a memory register $\vert m \rangle$ of length $n$, an auxiliary register $\vert a \rangle$ of length $n$, and a control register $\vert u \rangle$ of length $2$ initialised as $\vert 01 \rangle$. $\vert m \rangle$ and $\vert a \rangle$ are initialised as $\vert m \rangle = \vert a \rangle = \vert 0\rangle^{\otimes n}$, with the full quantum register initialised as 
\begin{align}
    \vert \psi_0 \rangle &= \vert a \rangle \vert u \rangle \vert m \rangle \nonumber \\
     &= \vert 0_1 \dots  0_n \rangle \vert 0 1 \rangle \vert 0_1\dots  0_n \rangle .
\end{align}

Each of the binary vectors are encoded sequentially. For each iteration of the encoding algorithm, a new state is generated in the superposition (excluding the final iteration). The new state generated is termed as the \textit{active} state of the next iteration. All other states are said to be \textit{inactive}. Note, in each iteration of the algorithm, the active state will always be selected with $\vert u \rangle = \vert 01\rangle$. 

During a single iteration, a binary vector is stored in integer format, which is then serially encoded bit-wise into the auxiliary register $\vert a \rangle$ resulting in the state $\vert \psi_1 \rangle$:
\begin{align}
    \vert \psi_1 \rangle & =  \vert a_1^{(1)}\dots a_n^{(1)}\rangle\vert 01\rangle\vert0_1\dots0_n \rangle.
\end{align}

This binary representation is then copied into the memory register $\vert m \rangle$ of the active state by applying a $2\textrm{CX}$ gate on $\vert \psi_1 \rangle$:
\begin{align}
\label{eq:encode_psi2}
    \vert \psi_2 \rangle & =  \prod\limits_{j=1}^{n} \textrm{2CX}_{a_j^{(i)} u_2 m_j} \vert \psi_1 \rangle.
\end{align}
Next, we apply a $\textrm{CX}$ followed by a $\textrm{X}$ gate to all qubits in $\vert m \rangle$ using the corresponding qubits in $\vert a \rangle$ as controls:
\begin{align}
\label{eq:encode_psi3}
  \vert \psi_3 \rangle & =  \prod\limits_{j=1}^{n} \textrm{X}_{m_j} \textrm{CX}_{a_j^{(i)} m_j}  \vert \psi_2 \rangle .
\end{align}
This sets the qubits in $\vert m\rangle$ to $1$ if the respective qubit index in both $\vert m \rangle$ and $\vert a \rangle$ match, else to $0$. Thus, the state whose register $\vert m \rangle$ matches the pattern stored in $\vert a \rangle$ will be set to all $1$'s while the other states will have at least one occurrence of $0$ in $\vert m \rangle$.

Now that the state being encoded has been selected, an $\textrm{nCX}$ operation is applied to the first qubit in the auxiliary register using the qubits in $\vert m \rangle$ as the controls:
\begin{align}
\label{eq:encode_psi4}
	\vert \psi_4 \rangle & =  \textrm{nCX}_{m_1\dots m_n u_1}  \vert \psi_3 \rangle.
\end{align}

\noindent The target qubit whose initial value is $0$ will be set to $1$ if $\vert m \rangle$ consists of only $1$'s. This is the case when the pattern in $\vert m \rangle$ is identical to the pattern being encoded (i.e. the pattern stored in $\vert a \rangle$).

In order to populate a new state into the superposition, it is required to effectively `carve-off' some amplitude from the existing states so the new state has a non-zero coefficient. To do this, we apply a controlled unitary matrix $\textrm{CS}^{(i)}$ to the second auxiliary qubit $u_2$ using the first auxiliary qubit $u_1$ as a control:
\begin{align}
\vert \psi_5 \rangle  =  \textrm{CS}_{u_1 u_2}^{(p+1-i)}  \vert \psi_4 \rangle,
\end{align}
\noindent where
\begin{align}
\textrm{S}^{(i)} &= \begin{bmatrix}
\sqrt{\frac{i-1}{i}} & \frac{1}{\sqrt{i}} \\
-\frac{1}{\sqrt{i}}  & \sqrt{\frac{i-1}{i}}
\end{bmatrix} \nonumber\\
&=  R_y\left(\phi(i)\right),
\end{align}
with $i \in \mathbb{Z^{+}}$, and $\phi(i) = - \cos^{-1}\left((i-2)/i\right) $. The newly generated state will be selected with $\vert u\rangle = \vert 11 \rangle$, while the previous active state used to `carve-off' this new state selected with $\vert u\rangle = \vert 10 \rangle$. All other states will be selected with $\vert u\rangle = \vert 00 \rangle$. 

To apply the next iteration of the algorithm we uncompute the steps from equations \eqref{eq:encode_psi2} - \eqref{eq:encode_psi4} as:
\begin{align}
	\vert \psi_6 \rangle & =  \textrm{nCX}_{m_1\dots m_n u_1}  \vert \psi_5 \rangle, \\
	\vert \psi_7 \rangle & =  \prod\limits_{j=n}^{1} \textrm{CX}_{a_j^{(i)} m_j} \textrm{X}_{m_j}  \vert \psi_6 \rangle, \\
	\vert \psi_8 \rangle & =  \prod\limits_{j=n}^{1} 2\textrm{CX}_{a_j^{(i)} u_2 m_j} \vert \psi_7 \rangle.
\end{align}
This results in the previous active state now being selected with $\vert u \rangle = \vert 00 \rangle$ while the new state with $\vert u \rangle = \vert 01 \rangle$, which identifies it as the new active state. The previous active state's memory register now contains the pattern $\{ a_1^{(i)}, \dots, a_n^{(i)} \}$ while the new active state's memory register is set to all zeroes. 

Finally, the register $\vert a \rangle$ for every state must be set to all zeroes by sequentially applying \textrm{X} gates to each qubit in $\vert a \rangle$ according to the pattern that was just encoded. The quantum register is now ready for the next iteration to encode another pattern. Following the encoding of all patterns, our state will be 
\begin{align}\label{eqn:encoded_memory}
    \vert \psi \rangle &= \vert a \rangle \vert u \rangle \vert m \rangle \nonumber \\
    &= \vert 0_{1}\dots 0_{n} \rangle\vert 00 \rangle \left( \frac{1}{\sqrt{N}}\sum\limits_{i=1}^{N} \vert p_1^{(i)}\dots p_n^{(i)} \rangle \right).
\end{align}
Note, this algorithm assumes that the number of patterns to be encoded is known beforehand, which is required to generate the set of $S^i$ matrices and apply them in the correct order. The total number of qubits used in this algorithm is $2n+2$, of which $n+2$ are reusable after the implementation since the qubits in $\vert a \rangle$ and $\vert u \rangle$ are all reset to $\vert 0\rangle$ upon completion.

The additional $n+2$ qubits allows for them to be used as intermediate scratch to enable the large n-controlled operations during the encoding stages. This ensures that we can perform the $\textrm{nCX}$ operations with a linear, rather than polynomial, number of two-qubit gate calls~\cite{barenco1995}.


\subsection{Representing patterns using encoded data}
\label{sec:compare_test_encoded}

The purpose of this methodology is to represent a single test pattern using the previously encoded meaning-space. The relative distance between each meaning-space state pattern and the single test pattern $x = \{x_1,\dots, x_n\}$ is then encoded into the amplitude of each respective meaning-space state pattern. Thus, each represented state will have a coefficient proportional to the Hamming distance between itself and the test pattern. The method we present below calculates the binary difference between the target state's bit-string and the test pattern, denoted by $d_H$. 

The algorithm assumes that we already have $N$ states of length $n$ encoded into the memory register $\vert m \rangle$. The subsequent encoding requires $2n+1$ qubits; $n$ qubits to store the test pattern, a single qubit register which the rotations will act on, and $n$ qubits for the memory register. As our previously used encoding stage required $2n+2$ qubits, we can repurpose the $\vert a \rangle$ and $\vert u \rangle$ registers as the test pattern and rotation registers respectively. Our meaning-space patterns are encoded in the memory register $\vert m \rangle$, with registers $\vert a \rangle$ and $\vert u \rangle$ initialised as all $0$'s. Hence, our initial state is given by eq.~\eqref{eqn:encoded_memory}.

Next, the test pattern $x = \{x_1,\dots,x_n\}$ is encoded into the register $\vert a \rangle$ sequentially by applying a \textrm{X} gate to each qubit whose corresponding classical bit $x_i$ is set:

\begin{align}\label{eq:encoded_test}
    \vert \psi_1^{\prime} \rangle = \vert x_{1}\dots x_{n} \rangle\vert 00 \rangle \left(\frac{1}{\sqrt{N}}\sum\limits_{i=1}^{N} \vert p_1^{(i)}\dots p_n^{(i)} \rangle \right).
\end{align}

Rather than overwriting register $\vert a\rangle$ with the differing bit-values, a two qubit controlled $R_y(\theta)$ ($2\textrm{C}R_y$) gate is applied, such that $ \theta = \frac{\pi}{n}$. This is done by iteratively applying the 2-controlled $R_y$ gate with $a_j$ and $m_j$ as control qubits to rotate $\vert u \rangle$ if both control qubits are set for $j = 1,\dots,n$. The operation is performed twice, such that $a_j=1, m_j=1$ and by appropriately flipping the bits prior to use for $a_j=0, m_j=0$.

Finally, the test pattern stored in register $\vert a \rangle$ is reset to consist of all $0$'s by applying a \textrm{X} gate to each qubit in $\vert a \rangle$ whose corresponding classical bit is set to $1$.

The above process can be written as follows:

\begin{multline}
    \label{eq:hamming_complete_alg}
    \vert \psi_2^{\prime} \rangle = \prod\limits_{j=1}^{n} \textrm{X}_{a_j}\textrm{X}_{m_j}  2\textrm{CR}_{y}^{(a_j,m_j,u_1)} \\ \textrm{X}_{a_j}\textrm{X}_{m_j} 2\textrm{CR}_{y}^{(a_j,m_j,u_1)} \vert \psi_1 \rangle,
\end{multline}
where the state after application is given by
\begin{equation}
    \vert \psi^{\prime} \rangle = \vert 0 \rangle^{\otimes (n+1)}  \frac{1}{\sqrt{k}}\displaystyle\sum\limits_{j=1}^{k}  \left[ \cos\left( \phi_j \right)\vert 0 \rangle + \sin\left(\phi_j\right)\vert 1 \rangle \right] \otimes \vert p^{(j)} \rangle,
\end{equation}
with
\begin{equation}
\phi_j = \frac{d_H(p^{(j)}, x)\pi}{n} = \frac{\pi}{n} \displaystyle\sum\limits^{n}_{l=1} p^{(j)}_l \oplus x_l.
\end{equation} 
Applying the linear map 
\begin{equation}\label{eqn:proj_op}
P  = \mathds{1}^{\otimes (n+1)} \otimes \vert 1 \rangle \langle 1 \vert \otimes \mathds{1}^{\otimes n},
\end{equation}
we represent the meaning-space states weighted by the Hamming distance with the test pattern, $x$. The state following this is given by
\begin{equation}\label{eq:post_hamming}
    \vert \psi^{\prime}_{x} \rangle = \frac{1}{\sqrt{k \langle \psi^{\prime} \vert P \vert \psi^{\prime} \rangle}}\displaystyle\sum\limits_{j}^{k} \sin(\phi_j) \vert p_j \rangle,
\end{equation}
where qubit registers $\vert a \rangle=\vert 0 \rangle ^{\otimes n}$ and $\vert u \rangle = \vert 01 \rangle$ are left out for brevity.


With the above method we can examine the similarity between patterns mediated via the meaning space. While one may directly calculate the Hamming distance between both register states as a measure of similarity, by doing this we lose distributional meaning discussed from Section~\ref{subsec:basis_words}. As such, we aim to represent both patterns in the meaning-space, and examine their resulting similarity using state overlap, with the result defined by
\begin{align}
    F(x^{(0)},&x^{(1)}) = \nonumber \\ 
     &\left\vert \frac{1}{k\sqrt{\langle P^{(0)} \rangle\langle P^{(1)}  \rangle}} \displaystyle\sum\limits_{j=1}^{k}\sin\left(\phi_{j}^{(0)}\right)\sin\left(\phi_{j}^{(1)}\right) \right\vert^2,
    \label{eqn:state_overlap}
\end{align}
with $\langle P^{(i)} \rangle = \langle x^{(i)} \vert P \vert x^{(i)} \rangle$, and $x^{(i)}$ as test pattern $i$.

\section{\label{sec:results}Results}

\subsection{Small-scale example}\label{subsec:example_hamming}
We now demonstrate an example of the method outlined in Sec.~\ref{sec:methods} for a sample representation and sentence comparison problem.

We opt for the simplified \texttt{noun-verb-noun} sentence structure, and define sets of words within each of these spaces, through which we can construct our full meaning space, following an approach outlined in~\cite{Coecke_Sadrzadeh_Clark_2010}. For nouns, we have: (i) subjects, $n_s = \{\textsc{adult},\textsc{child},\textsc{smith},\textsc{surgeon}\}$; and (ii) objects, $n_o = \{\textsc{outside},\textsc{inside}\}$. For verbs, we have $v = \{\textsc{stand},\textsc{sit},\textsc{move},\textsc{sleep}\}$. With these sets, we can represent the full meaning-space as given by
\begin{equation}\label{eqn:john_mary}
\left(
\begin{array}{c}
\textsc{adult} \\\\
\textsc{child} \\\\
\textsc{smith} \\\\
\textsc{surgeon} \\\\
\end{array}
\right) \otimes
\left(
\begin{array}{c}
\textsc{stand} \\\\
\textsc{sit} \\\\
\textsc{move} \\\\
\textsc{sleep} \\\\
\end{array}
\right) \otimes
\left(
\begin{array}{c}
\textsc{outside} \\\\
\textsc{inside} \\\\
\end{array}
\right).
\end{equation}

Whilst all combinations may exist, subjected to a given training corpus, only certain patterns will be observed, allowing us to restrict the information in our meaning-space. For simplicity, we can choose our corpus to be a simple set of sentences: $\textsc{John rests inside. Mary walks outside}$. To represent these sentences using the bases given by eq.~\eqref{eqn:john_mary}, we must determine a mapping between each token in the sentences to the bases. In this instance, we manually define the mapping by taking the following meanings:
\begin{itemize}
    \item $\textsc{John is an adult, and a smith}$. The state is then given as: \\ $\vert \textsc{John} \rangle = 1/\sqrt{2}\left(\vert \textsc{adult} \rangle + \vert \textsc{smith} \rangle\right)$, which is a superposition of the number of matched entities from the basis set.
    \item $\textsc{Mary is a child, and a surgeon}$. Similarly, the state is given as: \\ $\vert \textsc{Mary} \rangle = 1/\sqrt{2}\left(\vert \textsc{child} \rangle + \vert \textsc{surgeon} \rangle\right)$, following the same procedure as above.
\end{itemize}

We also require meanings for \textsc{rests} and \textsc{walks}. If we examine synonyms for \textsc{rests} and cross-compare with our chosen vocabulary, we can find \textsc{sit} and \textsc{sleep}. Similarly, for \textsc{walks} we can have \textsc{stand} and \textsc{move}. We can define the states of these words as $\vert \textsc{rest} \rangle = 1/\sqrt{2}\left(\vert \textsc{sit} \rangle + \vert \textsc{sleep} \rangle\right)$ and $\vert \textsc{walk} \rangle = 1/\sqrt{2}\left(\vert \textsc{stand} \rangle + \vert \textsc{move} \rangle\right)$.
Now that we have a means to define the states in terms of our vocabulary, we can begin constructing states to encode the data.

We begin by tokenising the respective sentences into the 3 different categories: subject nouns, verbs, and object nouns. With the sentence tokenised, we next represent them as binary integers, and encode them using the processes of Sec.~\ref{sec:methods}. The basis tokens are defined in table~\ref{tbl:basis_encoding}.
\begin{table}[h!]
    \centering
    \rowcolors{2}{gray!20!white}{gray!5!white}
    \begin{tabular}{ |c|C{0.25\columnwidth}|C{0.25\columnwidth}| }
        \textbf{Dataset} & \textbf{Token} & \textbf{Bin.~Index}\\
        \hline
${n_s}$ & {adult} & 00 \\
${n_s}$ & {child} & 11 \\
${n_s}$ & {smith} & 10 \\
${n_s}$ & {surgeon} & 01 \\
        \hline
${v}$ & {stand} & 00 \\
${v}$ & {move} & 01 \\
${v}$ & {sit} & 11 \\
${v}$ & {sleep} & 10 \\
        \hline
${n_o}$ & {inside} & 0 \\
${n_o}$ & {outside} & 1 \\
        \hline
    \end{tabular}
    \caption{Basis data}
    \label{tbl:basis_encoding}
\end{table}
We define the mapping of ``John rests inside, Mary walks outside'' to this basis in table~\ref{tbl:sentence_encoded}.
\begin{table}[h!]
    \centering
    \rowcolors{2}{gray!20!white}{gray!5!white}
    \begin{tabular}{ |c|C{0.25\columnwidth}|C{0.30\columnwidth}| }
        \textbf{Dataset} & \textbf{Token} & \textbf{State}\\
        \hline
        {$n_s$} & {John} & $(\vert 00 \rangle + \vert 10 \rangle)/\sqrt{2}$ \\
        {$n_s$} & {Mary} & $(\vert 01 \rangle + \vert 11 \rangle)/\sqrt{2}$ \\
        \hline
        {$v$} & {walk} & $(\vert 00 \rangle + \vert 01 \rangle)/\sqrt{2}$ \\
        {$v$} & {rest} & $(\vert 10 \rangle + \vert 11 \rangle)/\sqrt{2}$ \\
        \hline
        {$n_o$} & {inside} & $\vert 0 \rangle$  \\
        {$n_o$} & {outside} & $\vert 1 \rangle$  \\
        \hline
    \end{tabular}
    \caption{Sentence data encoding using basis from Table~\ref{tbl:basis_encoding}.}
    \label{tbl:sentence_encoded}
\end{table}

If we consider the \textsc{John} and \textsc{Mary} sentences separately for the moment, they are respectively given by the states $(1/2)\vert 0 \rangle \otimes (\vert 10 \rangle + \vert 11 \rangle)\otimes (\vert 00 \rangle + \vert 10 \rangle) $ for John, and $(1/2)\vert 1 \rangle \otimes (\vert 00 \rangle + \vert 01 \rangle)\otimes (\vert 01 \rangle + \vert 11 \rangle)$ for Mary. Note that we choose a little endian encoding schema, wherein the subject nouns are encoded to the right of the register and object nouns to the left. Tidying these states up yields 
\begin{equation*}\nonumber 
\begin{array}{cc}
\textsc{John rests inside} \rightarrow \vert J\rangle & \\
 = \frac{1}{2}(\vert 01100 \rangle + \vert 01000 \rangle +\vert 01110 \rangle +\vert 01010 \rangle), & \\
& \\
\textsc{Mary walks outside}  \rightarrow  \vert M\rangle & \\ 
=\frac{1}{2}(\vert 10011 \rangle + \vert 10111 \rangle +\vert 10001 \rangle +\vert 10101 \rangle),  & \\
\end{array}
\end{equation*}
where the full meaning is given by $\vert m \rangle = \frac{\vert J \rangle + \vert M \rangle}{\sqrt{2}}$, which is a superposition of the 8 unique encodings defined by our meaning-space and sentences. 

From here we will next encode a test state to be stored in register $\vert a \rangle$ for representation using the encoded meaning-space. We use the pattern denoted by ``\textsc{Adult(s) stand inside}'', which is encoded as $\vert a \rangle = \vert 00000 \rangle$. Constructing our full state in the format of eq.~\eqref{eq:encoded_test}, we get
\begin{eqnarray*}\nonumber 
\vert \psi \rangle = \frac{1}{2\sqrt{2}} \vert 00000 \rangle \otimes \vert 00 \rangle \otimes (\vert 01100 \rangle + \vert 01000 \rangle + \vert 01110 \rangle \\ + \vert 01010 \rangle + \vert 10011 \rangle + \vert 10111 \rangle + \vert 10001 \rangle + \vert 10101 \rangle \left. \right).
\end{eqnarray*}
By following the steps outlined in Sec.~\ref{sec:compare_test_encoded}, rotating a single qubit from the control register $\vert u \rangle$ based on the Hamming distance between both registers, and applying the map from eq.~\eqref{eqn:proj_op}, the state of register $\vert m \rangle$ encodes a representation of the test pattern in the amplitude of each unique meaning-space state.

\begin{figure}[h!]
    \includegraphics[width=0.99\columnwidth]{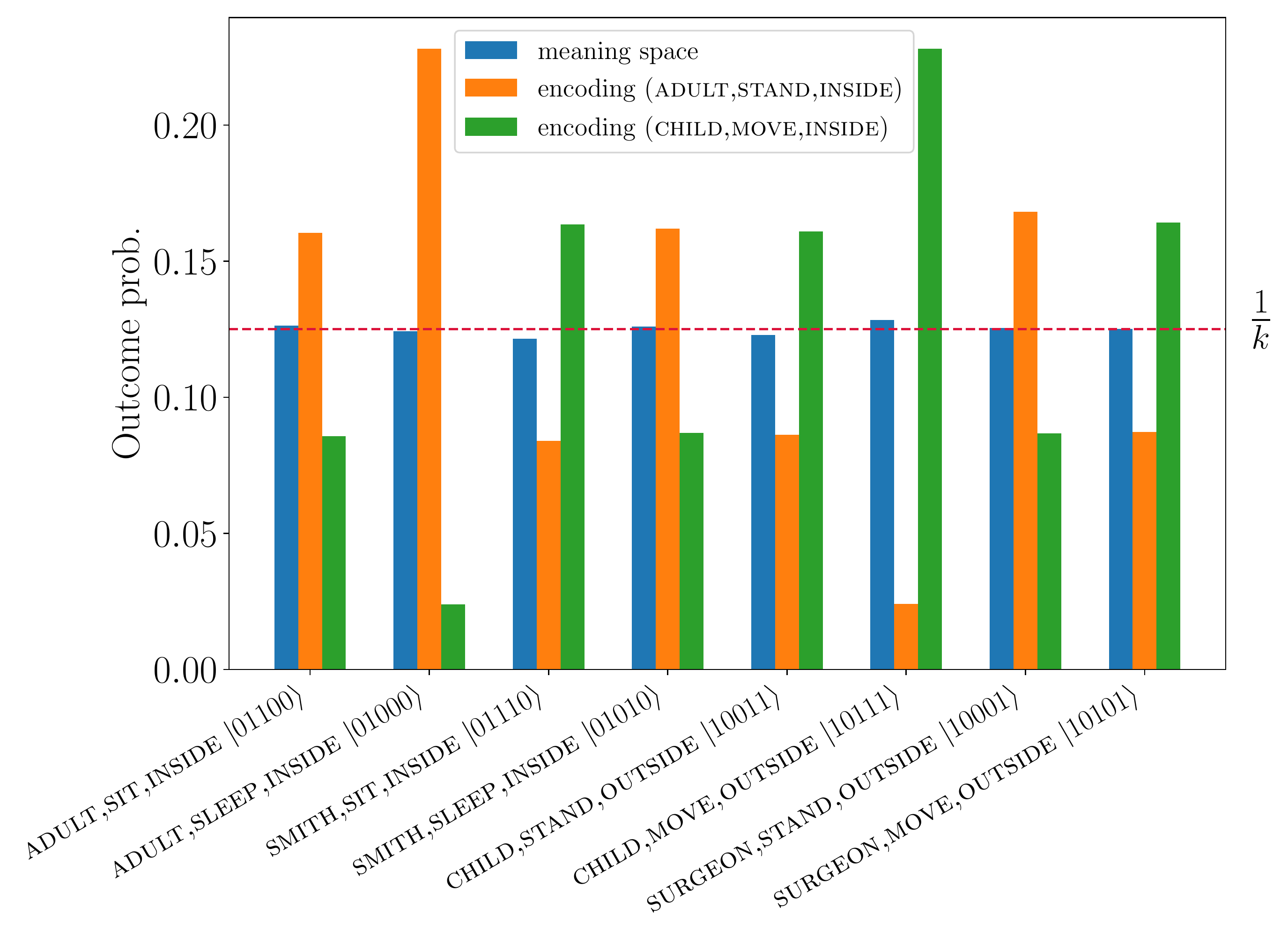}\\ 
  \caption{Sentence encoding state distribution taken by multi-shot preparation and measurement of $\vert m \rangle$ prior to, and post, the encoding of test patterns. Two distinct patterns are used: $00000\rightarrow \textsc{(adult, stand, inside)}$ (orange) and $00111\rightarrow \textsc{(child, move, inside)}$ (green). The distribution is sampled $5\times 10^4$ times, and shows how the Hamming distance weighting modifies the distribution relative to the $k=8$ unweighted meaning-space states (blue). }\label{fig:encoded_patterns}
\end{figure}

Through repeated preparation and measurement of the $\vert m \rangle$ register we can observe the patterns closest to the test. Figure~\ref{fig:encoded_patterns} shows the observed distribution using two different patterns; \textsc{adult, sit, inside} (00000, orange), and \textsc{child, move, inside} (00111, green) compared with the encoded meaning-space patterns following eq.~\eqref{eqn:encoded_memory} (blue).

Given this ability to represent patterns, we can extend this approach to examine the similarity of different patterns using eq.~\eqref{eqn:state_overlap}. One can create an additional memory register $\vert m^\prime \rangle$, and perform a series of SWAP tests between both encoded patterns, to determine a measure of similarity. For the above example, we obtain an overlap of $F(00000,00111) = 0.8602$, denoting a good degree of similarity, given our chosen meaning-space.

\subsection{Automated large-scale encoding}\label{subsec:example_hpc}
As that the previous example was artificially constructed to showcase the method, an automated workflow that determines the basis and mapped tokens, and performs the subsequent experiment is beneficial. Here we perform the same analysis, but using Lewis Carroll's ``Alice in Wonderland'' in an end-to-end simulation. 

\begin{figure}[htbp]
    \centering
    \includegraphics[width=0.95\columnwidth,trim={2.8cm 2.2cm 2.4cm 2.5cm},clip]{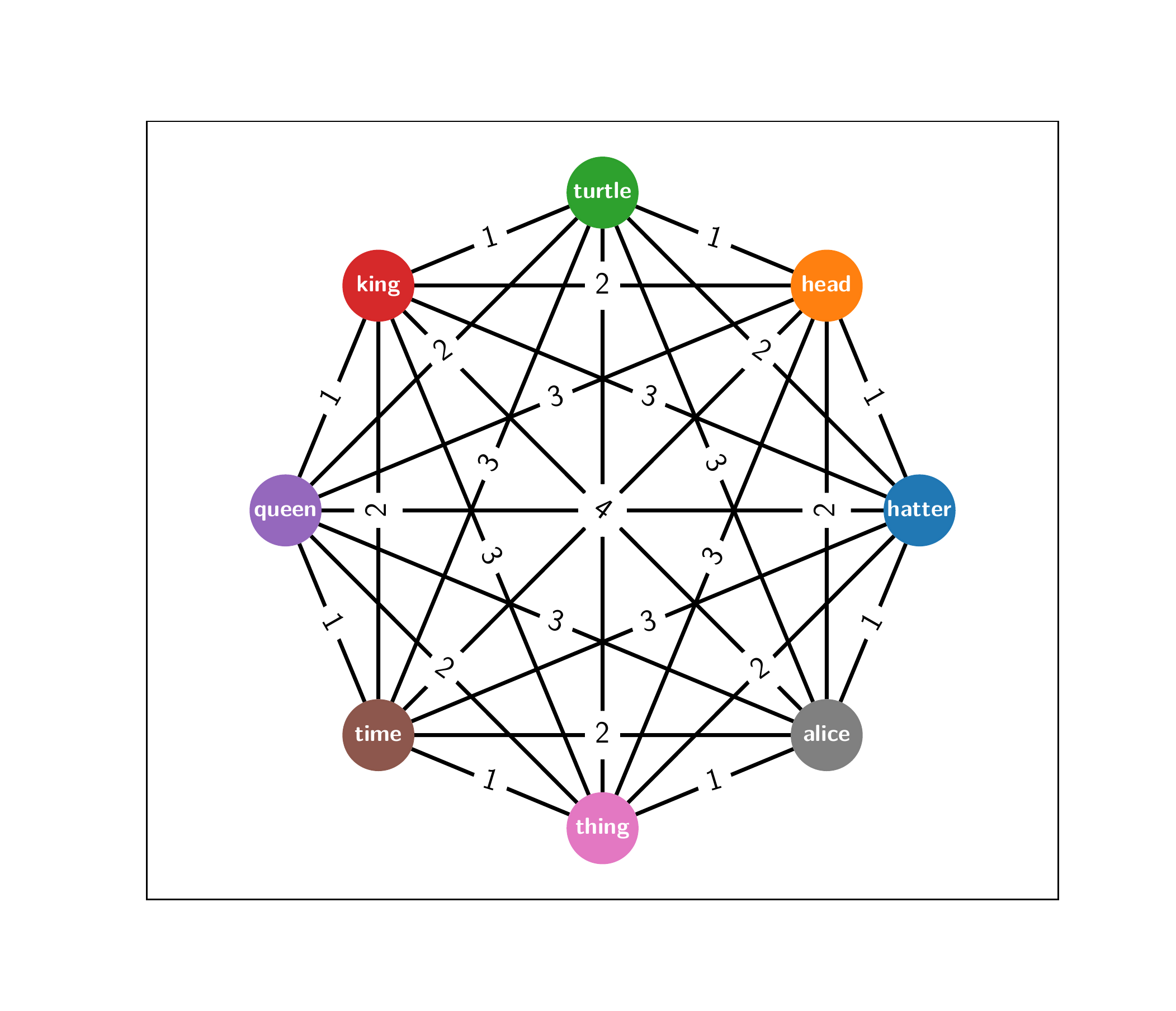}
    \caption{Relative ordering of an 8-basis set chosen for the noun dataset in ``Alice in Wonderland'', using the encodings and ordering given by Eq.~\eqref{eq:aiw_order}. The edge weight between each token shows the Hamming distance between the respective encoding patterns. }
    \label{fig:4q_aiw}
\end{figure}

To showcase the basis choice, we will consider the nouns basis set. We define a maximum basis set of 8 nouns ($N_{\textrm{nouns}}=8$), taken by their frequency of occurrence. Following the process outlined in Sec.~\ref{sec:methods}, we define a graph from these tokens, and use their inter-token distances to determine ordering following a minimum Hamiltonian cycle calculation. The resulting graph is shown by Fig.~\ref{fig:4q_aiw}. From here we map the tokens to an appropriate set of encoding bit-strings for quantum state representation, making use of eq.\eqref{eqn:cyclic_encoding}. The resulting set of mappings is :
\begin{equation}
\begin{array}{cccccc}
\textrm{head} & \rightarrow & \vert 0000\rangle, & \textrm{turtle} & \rightarrow &\vert 0001\rangle, \\
\textrm{hatter} & \rightarrow & \vert 0011 \rangle, & \textrm{king} & \rightarrow &\vert 0111\rangle, \\
\textrm{queen} & \rightarrow & \vert 1111 \rangle, &\textrm{time} & \rightarrow &\vert 1110\rangle, \\
\textrm{thing} & \rightarrow & \vert 1100\rangle, &\textrm{alice} & \rightarrow &\vert 1000\rangle.
\end{array}
\label{eq:aiw_order}
\end{equation}

We can now map the composite tokens onto the chosen basis encoding using a distance cut-off, $W_{\textrm{nouns}}$. Following the inter-word distance calculation approach used to determine basis order, we calculate the distance between the other corpus tokens and the respective basis set. Taking the set of all nouns in the corpus as $s_n$, and the noun basis set as $b_n \subset s_n $, for every token $t_n$ in $s_n$ we perform
\begin{align}
    t_n: s_n \mapsto b_n.
\end{align}
Tokens that fall outside $W_{\textrm{nouns}}$ are mapped to the empty set, $\emptyset$. This approach is then repeated for verbs, and lastly inter-dataset distances between noun-verb pairings, $W_{\textrm{nv}}$, which are used to discover viable sentences. The mapped composite tokens may then be used to create a compositional sentence structure by tensoring the respective token states. 

Following the previous example, we may examine the automatic encoding and representation of the string ``Hatter say queen'' to the meaning-space patterns. Given that representing the text in its entirety would be a substantial challenge, we limit the amount of information to be encoded by controlling the pre-processing steps as $N_\textrm{nouns} = 8$, $N_\textrm{verbs}=4$, $W_\textrm{nouns}=5$, $ W_\textrm{verbs}=5$ and $W_\textrm{vn}=4$. Here $N_\textrm{nouns}$ is again the number of basis nouns in both subject and object datasets, $N_\textrm{verbs}$ the number of basis verbs, $W_\textrm{nouns}$ and $W_\textrm{verbs}$ the cutoff distances for mapping other nouns and verbs in the corpus nouns to the basis tokens, and $W_\textrm{vn}$ is the cutoff distance to relate noun and verb tokens. 

\begin{figure}[h!]
    \centering
    \includegraphics[width=0.995\columnwidth,trim={0.cm 0.0cm 0cm 0cm 0cm},clip]{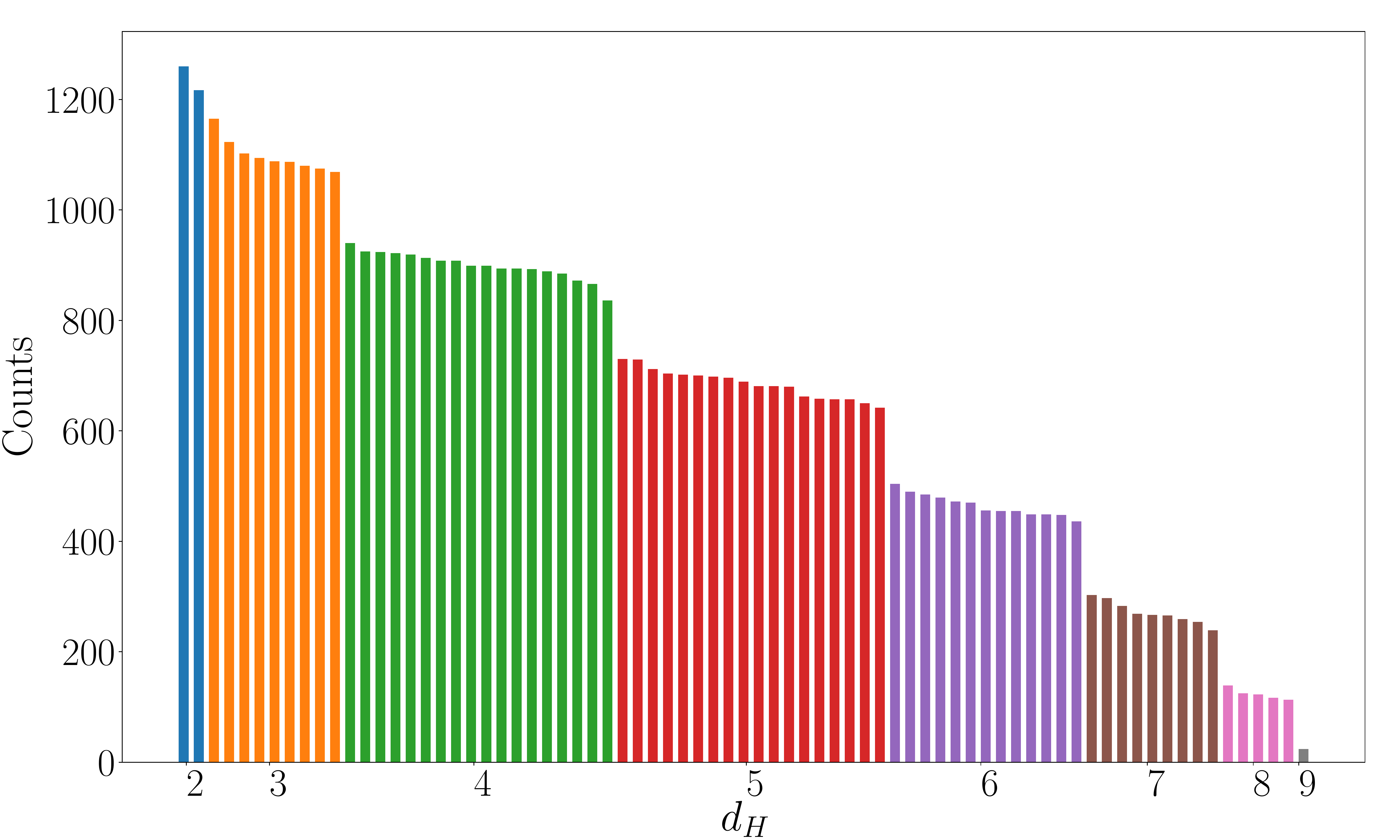}
    \caption{The result of measuring $\vert m \rangle$ states following the encoding of the string pattern ``Hatter say Queen''. This example uses a 75 unique pattern basis set, and taking $5\times 10^4$ samples to build the distribution. The Hamming distances of the labels are indicated on the $x$-axis, and differentiated by colour, where we can see clear distinction between the patterns in each Hamming category. The data mapping tokens to patterns in each category can be viewed in Appendix~\ref{app:app_sim_data}.}
    \label{fig:hatter_say_queen}
\end{figure}

For the above parameters, the method finds a subset of 75 unique patterns to represent the corpus. Following Section~\ref{subsec:example_hamming} one obtains the associated similarity of encoded elements by the resulting likelihood of occurrence, as indicated by Fig.~\ref{fig:hatter_say_queen}, where we have prepared and sampled the $\vert m \rangle$ register $5\times 10^4$ times to build the distribution. Clear step-wise distinctions can be observed between the different categories of Hamming-weighted states, with the full list presented in Appendix~\ref{app:app_sim_data} Table~\ref{tab:75_50k}. Given the basis encoding tokens from eq.~\eqref{eq:aiw_order}, the string ``Hatter say queen'' can be mapped to the value 995 (\verb+1111100011+ in binary).

As before, we can also compare patterns mediated via the meaning-space. For the pattern ``Hatter say Queen'', the most similar patterns are ``Hatter say King'' (\verb+0111100011+), ``Hatter go Queen'' (\verb+1111110011+) and ``Turtle say Queen'' (\verb+1111100001+) with overlaps of 0.974, 0.974 and 0.973 respectively. We include a variety of other encoded comparisons in the Appendix as Table.~\ref{tbl:overlap_patterns} to showcase the method.


\section{\label{sec:conclusions}Conclusions}
In this paper we have demonstrated methods for encoding corpus data as quantum states. Taking elements from the categorical distributional compositional semantic formalism, we developed a proof-of-concept workflow for preparing small and large scale data sets to be encoded, processed, and decoded using a given quantum register. We showed the preparation, encoding, comparison, and decoding of small and large datasets using the presented methods. 

Recent works have shown the importance of the reduction in classical data to be represented on a quantum system~\cite{2020arXiv200400026H}. The approach defined above follows an analogous procedure, representing the important elements of the corpus data using a  fundamental subset of the full corpus data. Using this subset, we have shown how to represent meanings, and subsequently the calculation of similarity between different meaning representations. We have additionally released all of this work as part of an Apache licensed open-source toolkit~\cite{QNLP_REPO}. 

For completeness, it is worth mentioning the circuit depths required to realise the above procedures. Taking the large scale example, we obtain single and two-qubit gate call counts of 2413 and 33175 respectively to encode the meaning space. This may be difficult to realise on current NISQ generation quantum systems, where the use of simulators instead allow us to make gains in understanding of applying these methods to real datasets.

The potential for circuit optimisation through the use of ZX calculus~\cite{CoeckeDuncan_ZX_2011}, or circuit compilation through tools such as CQC's $\textrm{t}\vert \textrm{ket} \rangle$ may offer more realistic circuit depths, especially when considering mapping to physical qubit register topologies~\cite{cowtan_et_al:LIPIcs:2019:10397}.

Very recent works on the implementation of the DisCoCat formalism on physical devices without the need for qRAM, have also emerged~\cite{discocat_medium_2020}. These methods may provide a more generalised approach to investigate quantum machine learning models in NLP and beyond, and have the potential to overcome the limitations discussed earlier with data encoding. We imagine the merging of this generalised approach~\cite{discopy} with the hybrid quantum-classical methods we have devised to allow interesting results and further development of this field. We leave this to future work.


\begin{acknowledgments}
We would like to thank Prof. Bob Coecke and Dr. Ross Duncan for discussions and suggestions during the early stages of this work. The work leading to this publication has received funding from Enterprise Ireland and the European Union's Regional Development Fund. The opinions, findings and conclusions or recommendations expressed in this material are those of the authors and neither Enterprise Ireland nor the European Union are liable for any use that may be made of information contained herein.  The authors also acknowledge funding and support from Intel during the duration of this project.

\end{acknowledgments}

\bibliographystyle{unsrt}
\bibliography{bibliography.bib}

\clearpage

\appendix

\section{Corpus Preparation}
\label{sec:Corpus_Preparation}
Our QNLP software solution~\cite{QNLP_REPO} can target most corpora provided that adequate pre-processing is conducted prior to the main routines of the application, and follows the outline approach from Sec.~\ref{sec:methods}. 

This approach has several variables that can be adjusted to control the operation of the pre-processing stage. The limiting number of top $N_\textrm{nouns}$ and $N_\textrm{verbs}$ are defined with the run-time parameters \verb+NUM_BASIS_NOUN+ and \verb+NUM_BASIS_VERB+, and defined as environment variables. The number of neighbouring nouns, $W_\textrm{nouns}$, and verbs, $W_\textrm{verbs}$, to consider when mapping the corpus tokens to basis tokens, are controlled by the run-time parameters \verb+BASIS_NOUN_DIST_CUTOFF+ and \verb+BASIS_VERB_DIST_CUTOFF+ respectively, and again defined as environment variables.

Finally, for forming \textit{noun-verb-noun} sentence structures, the number of neighbouring nouns to consider for determining the basis verbs, $W_\textrm{vn}$, are controlled through the environment variable \verb+VERB_NOUN_DIST_CUTOFF+. Additionally, the sentence is only valid if the inter-noun distance on a \textit{noun-verb-noun} structure is within $2W_\textrm{vn}$.
To choose appropriate values for these parameters, one must consider overall complexity of the corpus, number of \textit{noun-verb-noun} sentences, available qubit resources, and intended detail in representing the overall meaning. For the simplified example in Sec.~\ref{subsec:example_hamming}, we have a somewhat sparsely encoded set of patterns in the meaning space (8 patterns out of a possible 32), with a small number of qubits to represent the processing and assist with the encoding. A more complex text, with a larger basis set will require substantially more resources. For example, choosing \verb+NUM_BASIS_NOUN=10+ and \verb+NUM_BASIS_VERB=10+ using the discussed simplified cyclic encoding from eq.~\eqref{eqn:cyclic_encoding} will require at least 32 qubits. However, depending on the amount of information the pre-processing stage can extract, this may be an overestimate or underestimate of the required resources. 

\section{Software dependencies}
All results in this manuscript were generated using our QNLP toolkit, which is available at~\cite{QNLP_REPO}. Jupyter notebooks, packages and scripts exist for all operations described. We made use of the Intel Quantum Simulator~\cite{qhipster2020} to perform all quantum gate-level simulations, running on Kay, the Irish national supercomputer. To integrate our C++ work with Python we have made use of the pybind11 suite~\cite{pybind11}. All results obtained were through compilation with Intel® Parallel Studio XE 2019 Update 5 for distributed workloads (Sec.~\ref{subsec:example_hpc}), and GCC 9.2 for shared (OpenMP) workloads (Sec.~\ref{subsec:example_hamming}).

To analyse and prepare the corpus data for encoding into the quantum state-space, we have used the well-defined classical routines for corpus tokenisation and tagging from the NLTK~\cite{BirdKleinLoper09} and spaCy~\cite{spacy2} software suites.  For plotting we explicitly used pgfplots/tikz for Fig.~(1), and Matplotlib for all others~\cite{Hunter:2007}. We additionally used the Scipy ecosystem and pandas during results analysis and during the preprocessing stages\cite{numpy_array, 2020SciPy-NMeth, pandas_2011}. 

\clearpage 

\section{Encoded meaning-space data}\label{app:app_sim_data}
Table~\ref{tab:75_50k} is used to generate Fig.~\ref{fig:hatter_say_queen}. It encodes data from `Alice in Wonderland` using the preprocessing control parameters

\begin{itemize}
    \item Number of basis elements for state encoding: \verb+NUM_BASIS_NOUN=8 NUM_BASIS_VERB=4+
    \item Inter-token composite representation distance: \verb+BASIS_NOUN_DIST_CUTOFF=5,+
    \verb+BASIS_VERB_DIST_CUTOFF=5+
    \item Verb-noun distance cut-off for association: \verb+VERB_NOUN_DIST_CUTOFF=4+
\end{itemize}

\onecolumngrid
\begin{center}
\line(1,0){250}
\end{center}

\begin{table}[h!]
   \footnotesize
    \centering
    \rowcolors{2}{white}{gray!10!white}
\begin{minipage}[b]{.4\linewidth}\centering
\begin{tabular}{lccl}
\toprule
Label & Bin. pattern &  $d_H$ &   Count \\
\midrule
     king,go,queen &   1111110111 &    2 &  1260 \\
     king,say,time &   1110100111 &    2 &  1217 \\
    time,say,queen &   1111101110 &    3 &  1165 \\
      king,go,time &   1110110111 &    3 &  1123 \\
    queen,go,queen &   1111111111 &    3 &  1102 \\
  hatter,say,alice &   1000100011 &    3 &  1094 \\
   king,would,time &   1110000111 &    3 &  1088 \\
   king,say,hatter &   0011100111 &    3 &  1087 \\
      king,go,king &   0111110111 &    3 &  1080 \\
     head,go,queen &   1111110000 &    3 &  1075 \\
    queen,say,time &   1110101111 &    3 &  1069 \\
    alice,say,king &   0111101000 &    4 &   940 \\
     queen,go,king &   0111111111 &    4 &   925 \\
    alice,go,queen &   1111111000 &    4 &   924 \\
      head,go,king &   0111110000 &    4 &   922 \\
    king,go,hatter &   0011110111 &    4 &   919 \\
     time,go,queen &   1111111110 &    4 &   913 \\
 king,would,hatter &   0011000111 &    4 &   908 \\
  queen,would,time &   1110001111 &    4 &   908 \\
    alice,say,time &   1110101000 &    4 &   899 \\
  time,would,queen &   1111001110 &    4 &   899 \\
     time,say,king &   0111101110 &    4 &   894 \\
    king,say,alice &   1000100111 &    4 &   894 \\
     time,say,time &   1110101110 &    4 &   893 \\
  queen,say,hatter &   0011101111 &    4 &   889 \\
hatter,would,alice &   1000000011 &    4 &   885 \\
     queen,go,time &   1110111111 &    4 &   872 \\
   king,think,time &   1110010111 &    4 &   866 \\
   hatter,go,alice &   1000110011 &    4 &   836 \\
    thing,say,time &   1110101100 &    5 &   730 \\
 king,think,hatter &   0011010111 &    5 &   729 \\
   queen,say,alice &   1000101111 &    5 &   712 \\
  queen,think,time &   1110011111 &    5 &   704 \\
      time,go,time &   1110111110 &    5 &   702 \\
   time,would,king &   0111001110 &    5 &   700 \\
     alice,go,king &   0111111000 &    5 &   698 \\
   time,would,time &   1110001110 &    5 &   696 \\
     alice,go,time &   1110111000 &    5 &   689 \\
\bottomrule
\end{tabular}
\end{minipage}
\hspace{1cm}
\begin{minipage}[b]{.4\linewidth}\centering
\begin{tabular}{lccl}
\toprule
Label & Bin. pattern &  $d_H$ &   Count \\
\midrule
  alice,say,hatter &   0011101000 &    5 &   681 \\
hatter,think,alice &   1000010011 &    5 &   681 \\
   queen,go,hatter &   0011111111 &    5 &   680 \\
      time,go,king &   0111111110 &    5 &   662 \\
queen,would,hatter &   0011001111 &    5 &   658 \\
  king,would,alice &   1000000111 &    5 &   657 \\
    thing,go,queen &   1111111100 &    5 &   657 \\
     king,go,alice &   1000110111 &    5 &   650 \\
  alice,would,time &   1110001000 &    5 &   642 \\
queen,think,hatter &   0011011111 &    6 &   504 \\
  alice,think,time &   1110011000 &    6 &   490 \\
     head,go,alice &   1000110000 &    6 &   485 \\
  thing,would,time &   1110001100 &    6 &   479 \\
   alice,go,hatter &   0011111000 &    6 &   472 \\
 queen,would,alice &   1000001111 &    6 &   470 \\
   alice,say,alice &   1000101000 &    6 &   456 \\
    time,say,alice &   1000101110 &    6 &   455 \\
     thing,go,time &   1110111100 &    6 &   455 \\
  king,think,alice &   1000010111 &    6 &   449 \\
    queen,go,alice &   1000111111 &    6 &   449 \\
alice,would,hatter &   0011001000 &    6 &   448 \\
     thing,go,king &   0111111100 &    6 &   436 \\
  time,would,alice &   1000001110 &    7 &   303 \\
    alice,go,alice &   1000111000 &    7 &   297 \\
    alice,say,head &   0000101000 &    7 &   283 \\
     time,go,alice &   1000111110 &    7 &   269 \\
 alice,would,alice &   1000001000 &    7 &   267 \\
   thing,say,alice &   1000101100 &    7 &   266 \\
 queen,think,alice &   1000011111 &    7 &   259 \\
     time,say,head &   0000101110 &    7 &   254 \\
alice,think,hatter &   0011011000 &    7 &   239 \\
    thing,go,alice &   1000111100 &    8 &   139 \\
 thing,would,alice &   1000001100 &    8 &   125 \\
 alice,think,alice &   1000011000 &    8 &   123 \\
   time,would,head &   0000001110 &    8 &   117 \\
      time,go,head &   0000111110 &    8 &   113 \\
  alice,think,head &   0000011000 &    9 &    24 \\
    &     &      &      \\
\bottomrule
\end{tabular}
\end{minipage}
      \cprotect\caption{Results for Fig.~\ref{fig:4q_aiw} taking $5\times 10^4$ samples encoding AIW using the parameters \verb|NUM_BASIS_NOUN: 8|,
      \verb|NUM_BASIS_VERB: 4|, \verb|BASIS_NOUN_DIST_CUTOFF: 5|, 
      \verb|BASIS_VERB_DIST_CUTOFF: 5|,
      \verb|VERB_NOUN_DIST_CUTOFF: 4|, and comparing with the test pattern (`hatter,says,queen') with binary string $1111100011$. }
  \label{tab:75_50k}%
\end{table}

\twocolumngrid


\clearpage
\newpage
\mbox{~}
\clearpage
\newpage

\section{Overlap comparison data}\label{app:overlap_data}
Table~\ref{tbl:overlap_patterns} presents comparison data for the basis-token composed sentence ``Hatter say Queen'' and and a variety of other allowed sentence structures. Data  is again encoded from `Alice in Wonderland` using the preprocessing control parameters

\begin{itemize}
    \item Number of basis elements for state encoding: \verb+NUM_BASIS_NOUN=8 NUM_BASIS_VERB=4+
    \item Inter-token composite representation distance: \verb+BASIS_NOUN_DIST_CUTOFF=5,+
    \verb+BASIS_VERB_DIST_CUTOFF=5+
    \item Verb-noun distance cut-off for association: \verb+VERB_NOUN_DIST_CUTOFF=4+.
\end{itemize}

\onecolumngrid

\begin{center}
\line(1,0){250}
\end{center}

\begin{table}[h!]
   \footnotesize
    \centering
    \rowcolors{2}{white}{gray!10!white}
\begin{minipage}[b]{.46\linewidth}\centering
\begin{tabular}{m{3cm}m{2cm}m{1.8cm}}
\toprule
 Test pattern &  Bin. pattern &  Overlap \\
\midrule
     (hatter, go, queen) &   1111110011 &  0.974595 \\
     (hatter, say, king) &   0111100011 &  0.974003 \\
    (turtle, say, queen) &   1111100001 &  0.973813 \\
  (hatter, would, queen) &   1111000011 &  0.973788 \\
      (hatter, go, time) &   1110110011 &  0.959719 \\
  (hatter, think, queen) &   1111010011 &  0.955626 \\
     (turtle, say, time) &   1110100001 &  0.950021 \\
  (turtle, would, queen) &   1111000001 &  0.947137 \\
   (hatter, would, time) &   1110000011 &  0.945508 \\
      (hatter, go, king) &   0111110011 &  0.944756 \\
     (turtle, say, king) &   0111100001 &  0.943421 \\
   (hatter, would, king) &   0111000011 &  0.943293 \\
     (turtle, go, queen) &   1111110001 &  0.942680 \\
   (hatter, say, hatter) &   0011100011 &  0.942616 \\
   (hatter, think, time) &   1110010011 &  0.937847 \\
      (turtle, go, time) &   1110110001 &  0.927700 \\
  (turtle, think, queen) &   1111010001 &  0.923638 \\
   (hatter, think, king) &   0111010011 &  0.923143 \\
   (turtle, would, time) &   1110000001 &  0.918514 \\
      (head, say, queen) &   1111100000 &  0.918384 \\
    (hatter, go, hatter) &   0011110011 &  0.917483 \\
   (turtle, say, hatter) &   0011100001 &  0.913042 \\
   (turtle, would, king) &   0111000001 &  0.912276 \\
      (turtle, go, king) &   0111110001 &  0.907574 \\
   (turtle, think, time) &   1110010001 &  0.905863 \\
     (hatter, go, thing) &   1100110011 &  0.904857 \\
 (hatter, would, hatter) &   0011000011 &  0.904360 \\
       (head, say, time) &   1110100000 &  0.897983 \\
    (head, would, queen) &   1111000000 &  0.895655 \\
 (hatter, think, hatter) &   0011010011 &  0.891647 \\
   (turtle, think, king) &   0111010001 &  0.886336 \\
    (turtle, say, thing) &   1100100001 &  0.884995 \\
        (head, go, time) &   1110110000 &  0.884820 \\
    (turtle, go, hatter) &   0011110001 &  0.881117 \\
    (head, think, queen) &   1111010000 &  0.881114 \\
       (head, say, king) &   0111100000 &  0.880701 \\
  (hatter, would, thing) &   1100000011 &  0.874690 \\
  (hatter, think, thing) &   1100010011 &  0.874513 \\
 (turtle, would, hatter) &   0011000001 &  0.874446 \\
   (hatter, say, turtle) &   0001100011 &  0.872754 \\
     (head, would, time) &   1110000000 &  0.870438 \\
     (turtle, go, thing) &   1100110001 &  0.867190 \\
     (head, think, time) &   1110010000 &  0.864927 \\
     & & \\
\bottomrule
\end{tabular}
\end{minipage}
\hspace{1cm}
\begin{minipage}[b]{.46\linewidth}\centering
\begin{tabular}{m{3cm}m{2cm}m{1.8cm}}
\toprule
 Test pattern &  Bin. pattern &  Overlap \\
\midrule
     (head, say, hatter) &   0011100000 &  0.860190 \\
 (turtle, think, hatter) &   0011010001 &  0.856023 \\
     (head, would, king) &   0111000000 &  0.854352 \\
    (hatter, go, turtle) &   0001110011 &  0.851418 \\
     (hatter, say, head) &   0000100011 &  0.849715 \\
  (turtle, would, thing) &   1100000001 &  0.841915 \\
      (hatter, go, head) &   0000110011 &  0.841185 \\
      (head, go, hatter) &   0011110000 &  0.840234 \\
     (head, think, king) &   0111010000 &  0.838609 \\
   (turtle, say, turtle) &   0001100001 &  0.838448 \\
  (turtle, think, thing) &   1100010001 &  0.836987 \\
      (head, say, thing) &   1100100000 &  0.831755 \\
   (head, would, hatter) &   0011000000 &  0.827727 \\
    (turtle, say, alice) &   1000100001 &  0.824155 \\
       (head, go, thing) &   1100110000 &  0.823098 \\
 (hatter, would, turtle) &   0001000011 &  0.821690 \\
   (head, think, hatter) &   0011010000 &  0.818908 \\
 (hatter, think, turtle) &   0001010011 &  0.815651 \\
     (turtle, say, head) &   0000100001 &  0.815175 \\
     (turtle, go, alice) &   1000110001 &  0.814121 \\
    (turtle, go, turtle) &   0001110001 &  0.810484 \\
      (turtle, go, head) &   0000110001 &  0.801212 \\
   (hatter, think, head) &   0000010011 &  0.796473 \\
    (head, think, thing) &   1100010000 &  0.795634 \\
    (head, would, thing) &   1100000000 &  0.794096 \\
     (head, say, turtle) &   0001100000 &  0.790379 \\
      (head, say, alice) &   1000100000 &  0.789465 \\
   (hatter, would, head) &   0000000011 &  0.789114 \\
 (turtle, would, turtle) &   0001000001 &  0.786962 \\
       (head, say, head) &   0000100000 &  0.777728 \\
 (turtle, think, turtle) &   0001010001 &  0.775566 \\
      (head, go, turtle) &   0001110000 &  0.773961 \\
        (head, go, head) &   0000110000 &  0.772277 \\
  (turtle, think, alice) &   1000010001 &  0.770818 \\
  (turtle, would, alice) &   1000000001 &  0.766441 \\
   (turtle, think, head) &   0000010001 &  0.757195 \\
   (turtle, would, head) &   0000000001 &  0.754111 \\
    (head, think, alice) &   1000010000 &  0.748169 \\
   (head, would, turtle) &   0001000000 &  0.746573 \\
   (head, think, turtle) &   0001010000 &  0.743799 \\
    (head, would, alice) &   1000000000 &  0.739083 \\
     (head, think, head) &   0000010000 &  0.733556 \\
     (head, would, head) &   0000000000 &  0.725028 \\
          & & \\
\bottomrule
\end{tabular}
\end{minipage}
      \cprotect\caption{The overlap of representative encoding of test pattern (`hatter,says,queen') and a variety of other encodable patterns.}
  \label{tbl:overlap_patterns}%
\end{table}


\twocolumngrid

\end{document}